\documentclass[twocolumn,showpacs,preprintnumbers,amsmath,amssymb,nofootinbib,tighten,floatfix,prb]{revtex4}

\usepackage{color}
\usepackage{graphicx}
\usepackage{dcolumn} 
\usepackage{bm, bbm}      
\usepackage{curves}
\usepackage{epic}
\usepackage{wasysym}
\usepackage{epsf, times}
\usepackage{subfigure}
\usepackage{eufrak}
\usepackage{amsthm}
\usepackage[latin1]{inputenc}

\newcommand{\bc}{\begin{center}}
\newcommand{\ec}{\end{center}}

\def\ups{\uparrow}
\def\downs{\downarrow}
\def\a{\alpha}

\begin{document}
.
\title{Groundstatable fermionic wavefunctions and their associated
many-body Hamiltonians}

\author{Daniel Charrier$^{1,2}$}
\author{Claudio Chamon$^{1}$}

\affiliation{
$^1$ Physics Department, Boston University, Boston, MA 02215, USA
\\
$^2$ Laboratoire de Physique Th\'eorique, IRSAMC, UPS and CNRS, Universit\'e de Toulouse, F-31062 Toulouse, France
}

\date{\today}


\begin{abstract}
In the vast majority of many-body problems, it is the kinetic energy
part of the  Hamiltonian that is best known microscopically,
and it is the detailed form of the interactions between the particles,
the potential energy term, that is harder to determine from first
principles. An example is the case of high temperature
superconductors: while a tight-binding model captures the kinetic
term, it is not clear that there is superconductivity with only an
onsite repulsion and, thus, that the problem is accurately described
by the Hubbard model alone. Here we pose the question of whether, once
the kinetic energy is fixed, a candidate ground state is {\it
groundstatable or not}. The easiness to answer this question is strongly related to the presence or the absence 
of a sign problem in the system. When groundstatability is satisfied, it is simple to obtain the potential energy that
will lead to such a ground state. As a concrete case study, we apply
these ideas to different fermionic wavefunctions with superconductive or spin-density wave correlations and we also study the influence of Jastrow factors. The kinetic energy considered is a simple next nearest neighbor hopping term. 
\end{abstract}

\maketitle

\def\openone{\leavevmode\hbox{\small1\kern-4.2pt\normalsize1}}

\newcommand{\evec}{\mathbf e}
\newcommand{\kvec}{\mathbf k}
\newcommand{\svec}{\mathbf s}
\newcommand{\Rvec}{\mathbf R}
\newcommand{\bsigma}{\mathbf \sigma}
\newcommand{\ii}{{\rm i}}

\newcommand{\slapar}{\not \hskip -2 true pt \partial\hskip 2 true pt}
\newcommand{\slapartxt}{\not\!\!\!\!\partial}
\newcommand{\slaA}{\!\not\!\! A}
\newcommand{\slaAA}{\!\not\!\! A^{5}}
\newcommand{\slaM}{\ \backslash \hskip -8 true pt M}
\newcommand{\slaS}{\ \backslash \hskip -7 true pt \Sigma}

%
%

\section{Introduction} 

The problem of finding the ground state of a many-particle 
Hamiltonian is, in general, a daunting task. The problem is most
severe in the case of fermionic particles, where the infamous fermion
sign problem plagues solutions via numerical methods. In contrast, in
bosonic systems,  Monte Carlo methods are rather efficient in
simulating systems of reasonably large sizes. Certain methods, such as
Density Matrix Renormalization Group, avoid the sign problem, but are
mainly restricted to 1d or quasi-1d systems.

In this paper we step back from the problem of determining the ground
state of a given many-body Hamiltonian, and instead pose the following
question: fixing the kinetic energy part of the Hamiltonian, can a
{\it given} wavefunction be the ground state for {\it some} choice of
potential energy? Put simply, we ask if the wavefunction is {\it
groundstatable}. The question is trivial to answer for bosonic
system, as we discuss below, but in the case of fermions it is much
more difficult and subtle.

To illustrate this idea, let us start with a very simple example: the case of a single spin $1/2$ degree of freedom. 
Consider a Hamiltonian of the form:
\begin{equation}
\hat{H} = -\hat{\sigma}_x +\hat{V}_\alpha,
\end{equation}
where $\hat{\sigma}_x$ is the usual spin-flip operator and $\hat{V}_\alpha$ is 
a diagonal operator in the $\{ |\!\!\ups \rangle ,| \!\!\downs \rangle \}$ basis.
Here, we will not specify  $\hat{V}_\alpha$
  and try to diagonalize $\hat{H}$; instead , we will consider the wavefunction 
 \begin{equation}
 | \Psi_\alpha \rangle = \frac{1}{\sqrt{2(1+\alpha^2)}} \left[ (1-\alpha)\; |\!\! \ups \rangle + (1+\alpha) \; |\!\! \downs \rangle \right]
 \end{equation}
 and ask what the condition is on $\alpha$ so that $| \Psi_\alpha \rangle$ is the ground state of $\hat{H}$, for some a proper choice of $\hat{V}_\alpha$. To answer that, 
 the first step is to make $| \Psi_\alpha \rangle$ an eigenstate of $\hat{H}$ by imposing $\hat{H} | \Psi_\alpha \rangle = 0$ (so $| \Psi_\alpha \rangle$ is an eigenstate of $\hat{H}$ with energy zero). The expression of $\hat{V}_\alpha$ follows immediately and we can rewrite $\hat{H}$ in a matrix form as:
  \begin{equation}
    \hat{H} = 
      \left( 
        \begin{array}{cc}
          \!\!\! \frac{1+\alpha}{1-\alpha}& 
          -1 \\
          -1\;\;\;\; & 
           \frac{1-\alpha}{1+\alpha}
        \end{array}
      \right).
  \end{equation}
The two eigenvalues of this problem are $\lambda_1 = 0$ and $\lambda_2 = \frac{1+\alpha}{1-\alpha} + \frac{1-\alpha}{1+\alpha}$. Now, it is easy to see that $| \Psi_\alpha \rangle $ will be the ground state of $\hat{H}$ if and only if $\alpha < 1$ ({\it i.e.} if the wavefunction elements are all positive). We will say that $| \Psi_\alpha \rangle$ is \textit{groundstatable} for $\alpha < 1$. On the contrary, when $\alpha >1 $, $| \Psi_\alpha \rangle$ is an excited state of the problem and no longer groundstatable. At the boundary between the two cases, one component of the wavefunction vanishes at $\alpha = 1$. Then, the potential energy blows up and the eigenvalue $\lambda_2$ goes from $+\infty$ to $-\infty$. Of course, the property of groundstatability for a given wavefunction is directly related to the kinetic energy operator we have considered. Had we chosen a different operator, we would have reached a different conclusion on $\alpha$. The point is that once this operator is fixed, the problem is uniquely defined. 

This approach to the single spin Hamiltonian can be extended to a
many-body problem, where the kinetic energy is often chosen to be a
local hopping operator between nearby sites. Then, from the set of all
possible many-body wavefunctions, some are groundstatable and others
are not. It is of crucial importance to establish in which category a
given wavefunction belongs to, since it determines if this state is
allowed in nature. In general, for a given kinetic energy term, the
Hilbert space is broken down in regions in which the wavefunction
satisfies groundstatability; as we will see below, the level of
complexity of the partitions of the space into such regions is closely
related to the presence or the absence of a sign problem in the
Hamiltonian.

The paper is organized as follows. In section \ref{sec:problem} we
define the problem of groundstatability on a finite dimensional
Hilbert space and we show how it can be solved on a particular case
where the Hamiltonian admits a product form. Then, we present in
section \ref{sec:many-body} the main part of our work, namely how one
can build a Hamiltonian for which a given many-body fermionic
wavefunction is the groundstate. The numerical procedure is also
detailed in that section. Results are shown in section
\ref{sec:results}. We discuss the case of the wavefunction
for non-interacting fermions which allows us to illustrate the loss of
groundstatability in these systems. We then present the potentials
obtained from mean-field solutions of the Hubbard model, BCS
superconductors and spin-density waves (SDW).  These results are in
accordance with mean-field analysis. By considering additional Jastrow
factors, we also examine partially-projected BCS wavefunctions
relevant for the study of high-$T_c$ superconductors. Finally, a more
open problem, with a class of wavefunctions containing both
superconductivity and antiferromagnetism, is investigated.

\section{The problem}
\label{sec:problem}

\subsection{General considerations}

Let us consider a finite dimensional matrix example. Take a Hamiltonian matrix
$H_{C,C'}$, where $C,C'$ index the states in the $d_H$-dimensional
Hilbert space, for example the spatial configurations of fermions on a finite lattice.
Suppose that the off-diagonal elements $H_{C\ne C'}$
are known, and one wants to determine if the vector (state) with
components $\Psi_C$ can be the ground state if the matrix elements in
the diagonal are properly picked. There are two steps in the problem:
the first is trivial, to make $|\Psi\rangle$ an eigenstate, and the
second is the problem we pose, whether it can be {\it the} ground
state.

We start by determining the diagonal elements from the condition that
$|\Psi\rangle$ is an eigenstate. For simplicity, we shift again the
eigenvalue $\lambda_\Psi$ to zero, and solve for the $d_H$ variables
$H_{CC}$ in the diagonal:
\begin{equation}
\sum_{C'}\; H_{CC'}\;\Psi_{C'}=0\;
\Rightarrow
\;H_{CC}=-\sum_{C'\ne C}\;H_{CC'}\;\frac{\Psi_{C'}}{\Psi_C}
\;,
\label{eq:diagonal}
\end{equation}
so the Hamiltonian can be written as
\begin{eqnarray}
\hat H = -\frac{1}{2} \sum_{C\ne C'}
H_{CC'} \;\Big[
\frac{\Psi_{C'}}{\Psi_C} |C\rangle\langle C|
+\frac{\Psi_{C}}{\Psi_{C'}} |C'\rangle\langle C'|
\\
-|C\rangle\langle C'|-|C'\rangle\langle C|
\Big]
\;, \nonumber
\end{eqnarray}
which is a sum of projector operators acting on a 2-dimensional
subspace of states $C,C'$ (one can check that the operator within
brackets squares to a multiple of itself).

The problem of groundstatability is that it is not guaranteed, with
the Hamiltonian $H_{CC'}$ now determined, that $| \Psi \rangle$ is {\it the}
ground state, and not an excited state. What are the conditions on the
vector components $\Psi_C$ for it to be the ground state? If the
off-diagonal matrix elements are all non-positive, then one can make
use of the Perron-Frobenius theorem and the well-known connection to
stochastic dynamics~\cite{Castelnovo2005}, or alternatively cast the
Hamiltonian as a sum of positive semi-definite
projectors~\cite{Arovas-Girvin}. Basically, the condition of
groundstatability in this case is that $\Psi_C>0$, $\forall C$. This
is the case of matrix Hamiltonians for bosonic systems, and the
strictly negative or zero off-diagonal elements is related to the
absence of a sign-problem in the Hamiltonian. The problem of the
single spin $1/2$ mentioned in the introduction falls in this
category.  Now, one does not have the luxury of the stochastic mapping
to a problem with positive probabilities in general. If some
off-diagonal elements of the Hamiltonian are non positive, we lack any
general theorem to conclude on the groundstatability of the
wavefunction. Sometimes, it is possible to find a gauge transformation
to come back to the simpler case with all strictly negative or zero
off-diagonal elements; this happens for some spin models, like the
anti-ferromagnetic Heisenberg Hamiltonian, where the sign structure of
the ground state is given by the Marshall rule
~\cite{Marshall}. However, in fermionic problems and some frustrated
magnets, the problem remains unsolved.

\subsection{Hamiltonians with a separable form}

Before going to our main case of interest which is the fermionic
Hamiltonian with neighboring site hopping for kinetic energy term, we
would like to present another class of models where the question of
groundstatability can be completely and analytically answered even in
the absence of a Marshall-like rule. The problem of groundstatability
is defined for a given choice of kinetic energy operator, and we will
choose here the off-diagonal elements of the Hamiltonian $H^{(\pm)}_{C
\neq C'}=\pm w_C w_{C'}$ to be separable into products of $w_C,w_{C'}
\in \mathbb R$. Notice that these models are highly non-local
problems. However, they are interesting because they display the
fundamental difference between the presence/absence of a sign problem
in the system, mainly, the parameter space can be separated into
disconnected groundstatable regions in the $(+)$ case, whereas in the
$(-)$ case the groundstatable region is just made of a single block.

\begin{figure}
\includegraphics[scale=0.3]{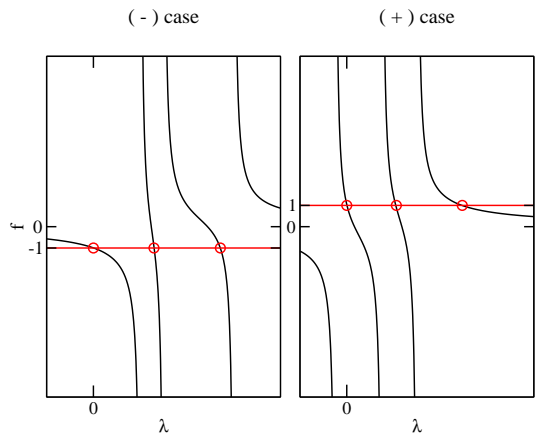} 
\caption{(color online) Solutions of the Schr\"odinger equation
(\ref{eq:pole-eq}) (circles) for a three dimensional Hilbert space
when $\lambda = 0$ is the ground state energy. Notice the positions of
the poles with respect to zero in the two cases.}
\label{fig:poles}
\end{figure}

Let us consider the two possibilities:   $H^{(+)}_{C\ne
C'}=+w_Cw_{C'}$ or $H^{(-)}_{C\ne
C'}=-w_Cw_{C'}$. Now, given a vector $\Psi_C$, we construct the
diagonal elements according to Eq.~(\ref{eq:diagonal}) so that
$\Psi_C$ is an eigenvector with eigenvalue zero. The Schr\"odinger equation for any eigenstate $| \psi^{\lambda} \rangle=\sum_C \psi^\lambda_C \;|C\rangle$ with energy $\lambda$ reads:
\begin{equation}
 \pm \sum_{C \neq C'} w_C w_{C'} \;\psi^\lambda_{C'} = \left( \lambda \pm \sum_{C \neq C'} w_C w_{C'} \frac{\Psi_{C'}}{\Psi_C}  \right)  \psi^\lambda_{C}
\end{equation}
and it is then straightforward to show that all eigenvalues $\lambda$ are solutions of the equation:
\begin{equation}
f_\pm(\lambda)=
\sum_C\frac{w_C^2}{\lambda\pm\left(\sum_{C'}w_{C'}\Psi_{C'}\right)w_C/\Psi_C}
=\pm 1.
\label{eq:pole-eq}
\end{equation}
$\lambda=0$ is indeed, by construction, a
solution. The state $|\Psi\rangle$ is the ground state if all other
solutions of Eq.~(\ref{eq:pole-eq}) are positive. It follows that
solutions of $f_-(\lambda)=-1$ satisfy $\lambda\ge 0$ if all poles of
the function $f_-(\lambda)$ are positive, and solutions of
$f_+(\lambda)=+1$ satisfy $\lambda\ge 0$ if one and only one of the
poles of the function $f_+(\lambda)$ is negative (see Figure \ref{fig:poles}).

Notice that in the ($-$) case, to fix the poles of $f_-(\lambda)$ to
be positive, the signs of groundstatable vectors $\Psi_C$ are related
to those of $w_C$, and one can thus write $\Psi_C={\rm
sgn}(w_C)\,|\Psi_C|$, which is a simple example of a Marshall sign. In
this case, the signs of the $w_C$'s can be gauged away from the
Hamiltonian, bringing it to the form that satisfy the Perron-Frobenius
theorem: $H_{C\ne C'}\to \tilde H_{C\ne C'}=-|w_C||w_C'|$ and
$\Psi_C\to\tilde\Psi_C= |\Psi_C|>0$.

\begin{figure}
\includegraphics[angle=0,scale=0.35]{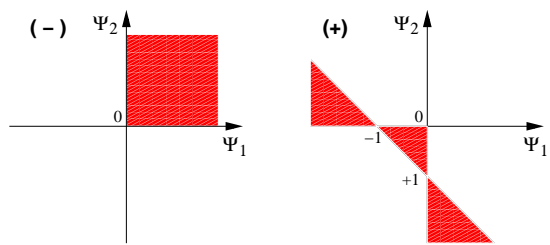} 
\caption{(color online) Examples of the domains where a wavefunction
$| \Psi \rangle =(\Psi_1,\Psi_2,\Psi_3)$ is groundstatable: we fix $\Psi_3=1$, and show
the regions in the $(\Psi_1,\Psi_2)$ plane where the wavefunctions are
groundstatable. ($-$) $H^{(-)}_{C\ne C'}=-1$
(Frobenius case) and ($+$) $H^{(+)}_{C\ne C'}=+1$ (non-Frobenius case), for
$C,C'=1,2,3$.}
\label{fig:simplex}
\end{figure}

The condition for groundstatability in the ($+$) case is richer. The
condition that one and only one of the poles of the function
$f_+(\lambda)$ is negative leads to $d_H$ distinct sectors in the
Hilbert space, each sector corresponding to the choice of which of the
$d_H$ poles is selected to be the negative one. More explicitly, the
condition on the poles is equivalent, for $\Psi_C\ne 0$, to:
\begin{eqnarray}
 w_{\bar C}\Psi_{\bar C}/(\sum_{C'}w_{C'}\Psi_{C'})&>&0
\, \, \mathrm{for} \ \bar C \\
w_C\Psi_C/(\sum_{C'}w_{C'}\Psi_{C'})&<0& \, \, \mathrm{for} \, \,C\ne \bar C
\label{eq:inequal}
\end{eqnarray}
 Each of these inequalities split the Hilbert space into two pieces via a hyperplane, and the $d_H$ conditions lead
to a simplex, and the choices of the $\bar C$ to $d_H$ such simplexes (see Fig.~\ref{fig:simplex} for simple examples on $3\times 3$ Hamiltonians). 
In most of the cases, the edges of the simplexes correspond to the
vanishing of one of the $ \Psi_C$ in the wavefunction. Indeed, coming
from a groundstatable region, if one component $\Psi_C$ changes sign the
inequalities (\ref{eq:inequal}) are violated. On the edge, $\Psi_C =
0$, the associated diagonal element $H_{CC}$ is infinite [see
Eq. (\ref{eq:diagonal})] and one of the positive eigenvalues diverges
towards $+\infty$ and reappears at $-\infty$.  Another edge is given by the equation
$\sum w_C \Psi_C = 0$. In this case, the whole hamiltonian is reduced
to the projector $ \hat{H} = | w \rangle \langle w |$ (this
corresponds to the diagonal line on Fig. \ref{fig:simplex}) and there
are only two eigenvalues: $0$ and $1$. From this point of view,
boundaries between groundstatable and non-groundstatable either
corresponds to ill-defined Hamiltonians or to highly degenerate
problems, the first case being the most common one.

For a Hamiltonian with off-diagonal matrix elements that cannot be
separated into a product as in the example above, the situation is
more complicated. Nonetheless, one should expect that the feature we
encountered in the simple case study should remain: generically the
set of groundstatable wavefunctions within the Hilbert space is
largely fragmented into regions. In the case of the example, there are
order $d_H$, the dimension of the Hilbert space, regions. This is to
be contrasted to the case where there is no fermion sign problem,
where there is one single region. This fragmentation of the
groundstatable set should be a generic feature of systems with
fermionic sign problems. Moreover, we also expect the diagonal part of
the Hamiltonian to abruptly change form when it leaves a
groundstatable region and to become singular at the boundary (the
edges of the simplexes). These abrupt changes in the Hamiltonian can
be used as telltales that the wavefunction, as function of some
parameter, exits a groundstatable region.

It is important to point out that, within a given groundstatable
region, many phases of matter can exist. Order parameters computed
from a groundstatable wavefunction can be used to classify the
phases. The groundstatable regions thus do not delimit single phases;
what they do demarcate are the regions where a wavefunction can
possibly correspond to a state of matter, for a fixed kinetic energy
term in the Hamiltonian.

\section{Many-body fermionic Hamiltonians}
\label{sec:many-body}

Let us now turn into a more practical application, and show how one
can implement the procedure of finding the potential energy term for
which a given fermionic many-body state $|\Psi \rangle$ is the ground
state, given a kinetic energy. We will consider specifically the case
where the kinetic energy comes from a tight-binding hopping term on a
lattice, which is common in many strongly-correlated electronic
problems. We consider the case of fermions on a square lattice, and
detail below the numerical procedure used to evaluate the potential
energy of the Hamiltonian.

Shifting the ground state energy to zero, the Hamiltonian 
\begin{equation}
\hat{H}_{\Psi} = \Hat{V}_{\Psi} 
-t \!\sum_{<ij>} c^{\dag}_{i\sigma}c_{j\sigma} + H.c. 
\label{hamiltonian}
\end{equation}
that we seek should satisfy
\begin{subequations}
\begin{eqnarray}
&&\hat{H}_{\Psi} | \Psi \rangle  = 0 
\label{condition}
\\
&&\hat{H}_{\Psi} |\lambda_n \rangle = 
\epsilon_n |\lambda_n \rangle, \quad \epsilon_n \geq 0,
\end{eqnarray}
\end{subequations}
for all eigenstates $|\lambda_n \rangle$. The $i,\sigma$ label the
site and the spin of the fermions, respectively. (Hereafter we set the
energy scale $t=1$.) The potential $\hat{V}_{\Psi}$ depends only on
the fermionic occupation operators $n_{i\sigma}$ and is uniquely
determined by the condition Eq.~(\ref{condition}) provided
$|\Psi\rangle$ is groundstatable. We will focus here on the general form
of this potential, addressing the question of groundstatability for
more specific cases.

We treat this problem in the configuration basis $\{|C\rangle\}$,
where a basis element stands for a set of positions of the $2N$
fermions, say $\{ \mathbf{R}_l^{\ups}\}_{l=1,\dots,N}$ for the up
spins and $\{ \mathbf{R}_m^{\downs}\}_{m=1,\dots,N}$ for the down
spins. The anticommutation relations between fermionic operators also
require enumerating the fermions and keeping the same ordering for
each configuration. The action of the kinetic operator $\hat{T}$ on an
ordered configuration $C$ can be understood by introducing a
configuration $\tilde{C'}$ such that $H_{C'C} \equiv \langle C' |
\hat{T} | C \rangle = \langle C' | \tilde{C'} \rangle$, where $C'$ is
another ordered configuration differing from $C$ by the local hopping
of a single electron. If configuration $\tilde{C'}$ is correctly
ordered, $H_{CC'}$ is equal to $-1$, otherwise it is equal to
$+1$. Thus, to determine the sign of $\langle C' | \hat{T} | C
\rangle$ one has to consider the positions of all the fermions in $C$
and $C'$. This is the sign problem, which bedevils  Monte Carlo
simulations on the Hubbard model in dimensions higher than one.

In this study, the fermionic many-body wavefunctions will take the form:
\begin{equation}
|\Psi \rangle = \frac{1}{\sqrt{\cal N}}\sum_{C} J_C \; 
\mathrm{det}\,(\phi_C)\;|C\rangle,
\end{equation}
where $J_C$ is a Jastrow factor that depends on the fermion occupation
numbers in configuration $C$, and $\phi_C$ is a $N\times N$ matrix
with elements that depend on the position of the particles in
configuration $C$, $[\phi_C]_{lm}\equiv\varphi
\left(\mathbf{R}_l^{\ups}-\mathbf{R}_m^{\downs} \right)$, with
$\varphi$ a function characterizing the correlations between pairs of
fermions\cite{Gros,Bouchaud}. This form includes the wavefunction for non-interacting fermions, BCS
and spin density wave (SDW) wavefunctions and also partial 
projections of these states \cite{Gros,Bouchaud}.  In the
configuration basis, the potential $\hat{V}_{\Psi}$ reads:
\begin{equation}
\langle C| \hat{V}_{\Psi}|C\rangle \equiv V_C 
= 
-\sum_{C'\ne C} H_{CC'} 
\frac{J_{C'}}{J_C} 
\frac{\det(\phi_{C'})}{\det(\phi_{C})},
\label{potential}
\end{equation}
supposing there are no vanishing determinants. Only neighboring
configurations, defined so that $C$ and $C'$ differ by the hopping of
a single fermion, contribute to the sum. To evaluate the sum, rather
than determining $H_{CC'}$ for each pair of configuration $C$ and
$C'$, we can use the fact that $\det(\phi_{C'})=\langle
C|\tilde{C'} \rangle\times \det(\phi_{\tilde{C'}})$ to rewrite the
potential $V_C$ as a function of configurations $\tilde{C'}$:
\begin{equation}
V_C 
= 
-\sum_{\tilde{C'}}{}^{'} 
\frac{J_{\tilde{C'}}}{J_C} \frac{\det (\phi_{\tilde{C'}})}{\det (\phi_{C})},
\end{equation}
where the primed sum is over configurations $\tilde{C'}$ that differ
from $C$ by the hopping of a single electron. We then compute $V_C$ by
calculating all the ratios $\det (\phi_{\tilde{C'}})/\det(\phi_{\tilde{C}})$ which are easy to calculate since the matrices
$\phi_{\tilde{C'}}$ differ from $\phi_{C}$ by the modification of one
row or one column, depending if an up or down spin hopped,
respectively -- recall that $[\phi_C]_{lm}\equiv\varphi
\left(\mathbf{R}_l^{\ups}-\mathbf{R}_m^{\downs} \right)$. Notice that,
by working directly with the positions $\mathbf{R}_l^{\ups}$ and
$\mathbf{R}_m^{\downs}$, issues of orderings of configurations
disappear from the problem.

To evaluate the operator $\hat{V}_{\Psi}$, we first compute $V_C$ for a large number of configurations $\mathcal{N}$.
 The configurations are chosen according to their weight $| \Psi_C |^2$ via a Metropolis algorithm, using the inverse update
  method for fermionic Monte Carlo ~\cite{Ceperley}. 
Then, we search for the best two-body approximation to the interaction,
neglecting three-body and higher order terms:
\begin{align}
\hat{H}_\Psi &= \tilde{H} + {\cal O}(\hat{n}_i \hat{n}_j \hat{n}_k) + ..., \nonumber \\ 
\tilde{H} &= \hat{T}+ E + U\sum_i \hat{n}_{i\ups}\hat{n}_{i\downs}+\frac{1}{2}\sum_{i \neq j}V_{ij}\, \hat{n}_{i} \hat{n}_{j} 
\label{eq:two-body}
\end{align}
where $\hat{n}_i\equiv \hat{n}_{i\ups}+\hat{n}_{i\downs}$. (Notice that for fixed
particle number the onsite potential can be written with $V_{ii}=U$,
with a constant shift absorbed into $E$.) The coefficients $V_{ij}$
are evaluated through a linear least square method (LLS) and the best
approximated solution is the one minimizing the sum of the squared
residuals $S = \langle (\hat{H}_\Psi - \tilde{H})^2\rangle $. This
quantity can be related to the overlap $\delta$ between the
wavefunction $| \Psi \rangle $, ground state of $\hat{H}$, and the
ground state $| \tilde{\Psi} \rangle$ of $\tilde{H}$, as follows.
Using perturbation theory on $\delta H=\tilde{H}- \hat{H}_\Psi$, we can
write the (unormalized) ground state wavefunction of $\tilde{H}$ as
\begin{equation}
|\tilde \psi \rangle = |\Psi\rangle + \sum_{n \neq 0} \;|n\rangle \;\frac{\langle n | \left( \hat{H}_\Psi - \tilde{H} \right) | \Psi \rangle}{E_n}+\dots 
\end{equation}
where $ | n \rangle$ and $E_n$ are the eigenstates and eigenvalues of
$\hat{H}$. The norm of this state, $\langle \tilde \psi|\tilde \psi
\rangle$, can be related to the squared
residuals $S = \langle \Psi| (\hat{H}_\Psi - \tilde{H})^2|\Psi \rangle$:
\begin{eqnarray}
&&\langle \tilde \psi|\tilde \psi \rangle =
1+ \sum_{n \neq 0} \;\frac{\left|\langle n | \left( \hat{H}_\Psi - \tilde{H} \right) | \Psi \rangle\right|^2}{E_n^2}+\dots 
\nonumber\\
&&\le
1+ \frac{1}{E_1^2}\,\sum_{n \neq 0} \;\left|\langle n | \left( \hat{H}_\Psi - \tilde{H} \right) | \Psi \rangle\right|^2+\dots 
\nonumber\\
&&\le
1+ \frac{1}{E_1^2}\,\sum_{n} \;\langle \Psi| \left( \hat{H}_\Psi - \tilde{H} \right) |n \rangle \langle n| \left( \hat{H}_\Psi - \tilde{H} \right) | \Psi \rangle+\dots 
\nonumber\\
&&= 1+ S/E_1^2+\cdots
\end{eqnarray}
Overlapping $|\Psi\rangle$ with the normalized state $|\tilde{\Psi}
\rangle=|\tilde \psi \rangle\;\frac{1}{\sqrt{\langle \tilde
\psi|\tilde \psi \rangle}}$ yields (up to second order in perturbation
theory in $\delta H$)
\begin{equation}
\delta=|\langle \Psi|\tilde\Psi\rangle|^2\ge \frac{1}{1+ S/E_1^2}
\;.
\end{equation}
Therefore the squared residuals $S = \langle \Psi| (\hat{H}_\Psi -
\tilde{H})^2|\Psi \rangle$ which we obtain by approximating the
potential energy $V_\Psi$ by a two-body interaction are a measure of
the overlap between the ground states of $\hat H_\Psi$ and its
two-body approximation $\tilde H$. The smaller $S$, the closer the
overlap is to $1$. One can bound the overlap between the two
wavefunctions by noticing that, even if the system is gapless, the
excitation energy $E_1$ should be controlled by the finite size $L$ of
the system, and thus if $S$ is found to be much smaller than $E_1^2$,
the overlap will remain close $1$. In estimating the overlap
hereafter, we use the worst case scenario that the system is gapless,
with a wavevector $2\pi /L$ for the lowest energy excitation. (In the
computations of $\delta$ below, we assume linearly dispersing modes
with a velocity of order unity.)

We computed the potential and the associated overlap for several
wavefunctions. In each case, we considered a tilted lattice of size
$L^2 +1$ with odd $L$ and periodic boundary conditions to avoid
singularities of $d$-wave wavefunctions in reciprocal space
\cite{Randeria}.  10000 sweeps are usually considered for
equilibration.  Then, up to $\mathcal{N} = 80000$ configurations are
taken for the LLS method.  Computations have been made for various
system sizes ($L = 13,15,17,19$) at half filling ($N =
170,226,290,362$).  Because one can always change the constant $E$ by
a shift in all the coefficients $V_{ij}$, we add the additional
constraint $\sum_{|i-j| > R} V_{ij} = 0$ such that the last $L/2$
coefficients average to zero. We checked numerically that the results
do not depend on that specific choice of $R$.

We would like to emphasize here the difference between our method and
the traditional Variational Monte Carlo (VMC) method. Given a
wavefunction $ |\Psi \rangle$, the VMC method provides an upper bound
for the ground state energy of a Hamiltonian $\hat H$, {\it i.e.}, in
VMC the Hamiltonian is given, and a wavefunction is the target. By
varying the parameters contained in $|\Psi \rangle $, one can find the
best choice which minimizes the energy and then compute other operator
averages such as order parameters or correlations functions. However,
it is also possible that the real ground state of the system is so
different from $|\Psi \rangle $ that it cannot be reached by a
variation in the parameters. This systematic uncertainty is not
encountered in our approach as the Hamiltonian $\hat{H}_{\Psi}$
derives uniquely from $|\Psi \rangle$, {\it i.e.}, we inverted the
target to be the Hamiltonian and not the wavefunction. If $|\Psi
\rangle$ is groundstatable, it is the ground state of $\hat{H}_{\Psi}$
by construction.

\section{Results}
\label{sec:results}

\subsection{Fermions in 1D}

The potential obtained from the LLS expansion should give an adequate
description of the Hamiltonian when interactions are predominantly
two-body. Accordingly, when three-body and higher interactions are
totally absent, it should reproduce the exact form of the
Hamiltonian. Hence, to check the consistency of our method, we would
like to begin with a fermionic system with only two-body interactions
whose ground state is known exactly, and try to recover the
Hamiltonian starting from the wavefunction. Unfortunately, we lack any
exact results in two dimensions. So, we will preliminarily step back
to the one dimensional case where exact results are known, and
investigate systems of $N$ interacting fermions on a ring. An
appropriate case is the Hamiltonian with an inverse-square potential:
\begin{equation}
\hat{H} = -\sum_i \frac{\partial^2}{\partial x_i^2} +\frac{2\lambda (\lambda-1)\pi^2}{L^2}\sum_{i < j} \frac{1}{\sin^2\left(\frac{\pi (x_i-x_j)}{L}\right)}.
\label{eq:H-Sutherland}
\end{equation}
Depending on the value of $\lambda$, the potential can be either
attractive ($\lambda < 1$) or repulsive ($\lambda > 1$), $\lambda = 1$
corresponding to the non-interacting case. The ground state
wavefunction of this Hamiltonian has been found by Sutherland
\cite{Sutherland} to be of the product form
\begin{equation}
\Psi(x_1,...x_N)  = \prod_{j > k} {\sin^{\lambda}\left(\frac{\pi (x_j-x_k)}{L}\right)}
\;.
\label{eq:Sutherland}
\end{equation}

\begin{figure}
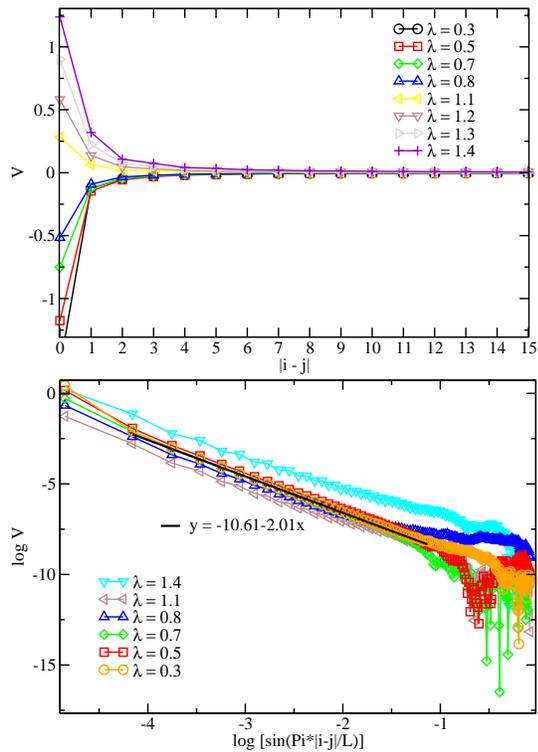

{\vspace*{-0.3cm}}\includegraphics*[width=7cm]{Calogero.eps}
\includegraphics*[width=7cm]{Calogerolog.eps} 
\caption {(color online)Top: Two-body potentials $V_{ij}$ as function
of the distance $|i-j|$ between sites for the Calogero-Sutherland
wavefunction. Bottom: $\log V$ as function of $\log \sin (\pi(x_i -
x_j)/L)$ for different values of $\lambda$.}
\label{fig:Sutherland}
\end{figure}

Let us now apply our procedure to the wavefunction
(\ref{eq:Sutherland}). The configuration basis is the set of positions
of $N$ spinless fermions on a ring of size $L$. For $\mathcal{N}$
configurations $|C \rangle$, we calculate $V_C$ using
Eq.~(\ref{eq:diagonal}) and then perform the LLS method. We do not
have to worry about groundstatability here, since the wavefunction is
always the groundstate for any $\lambda$. The potential energy we
obtain is presented in Fig.~\ref{fig:Sutherland} for different values
of $\lambda$, $L = 302$ and $\mathcal{N} = 80000$. We plot the set of
linear coefficients $\{U,V_{ij} \}$ as function of the distance
between sites $|i - j|$. As shown, we recover a potential falling
algebraically with the distance. The potential is attractive for
$\lambda < 1$, repulsive for $\lambda > 1$, and vanishes in between
(for $\lambda=1$). Indeed, the precise dependence of the potential
energy on the distance is found in quantitative agreement with that in
the Hamiltonian~\ref{eq:H-Sutherland}, with our method returning an
exponent $\gamma = 2.01 \pm 0.1$ for the power law decay shown in
Fig.~\ref{fig:Sutherland} (bottom panel). For each value of $\lambda$
studied, the overlap $\delta$ is found to be larger than $99.99 \%$.
 
\subsection{Fermions in 2D}

In two dimensions, things are less simple for two reasons: first,
except for very particular cases, we do not know if a wavefunction
will be groundstatable or if it will be an excited state. Secondly,
there are in general three-body and higher order terms in the
potential that makes the expansion (\ref{eq:two-body}) not exact and
that will result in a reduction of the overlap $\delta$. A good way to
proceed then is to start from wavefunctions we know \textit{a priori}
are groundstatable ({\it i.e.} by other means) and then adiabatically
deform them. By continuity, the resulting wavefunctions should also be
groundstatable unless a boundary is met. Moreover, as long as the
deviation of the overlap away from $\delta=1$ is small, the expansion
(\ref{eq:two-body}) should be relevant. How can we detect a boundary
then? Of course, we do not have any analytical criteria such as the
inequalities (\ref{eq:inequal}) here but we can make some basic
assumptions guided by what we learned from the separable case. At a
boundary, the diagonal part of the Hamiltonian is ill-defined so we
expect some kind of singularity in the Hamiltonian (notice that even
the simple spin $1/2$ example in the introduction displayed such
singular behavior as one crossed the boundary of
groundstatability). The singular behavior signaling the boundary of a
groundstatable region can appear in the set $\{U, V_{ij} \}$ or in the
overlap $\delta$. In a finite system, that means a strong dependance
of the Hamiltonian with the system size. Note the difference with a
phase transition, where it is the wavefunction which displays singular
behavior at a critical point as the Hamiltonian is smoothly varied;
here the problem is inverted, as it is the Hamiltonian that is
singular at the boundary of the groundstatable region as the
wavefunction is smoothly varied.

\subsubsection{The deformed non-interacting wavefunction}

\begin{figure}
{\vspace*{-0.3cm}}\includegraphics*[width=5.5cm]{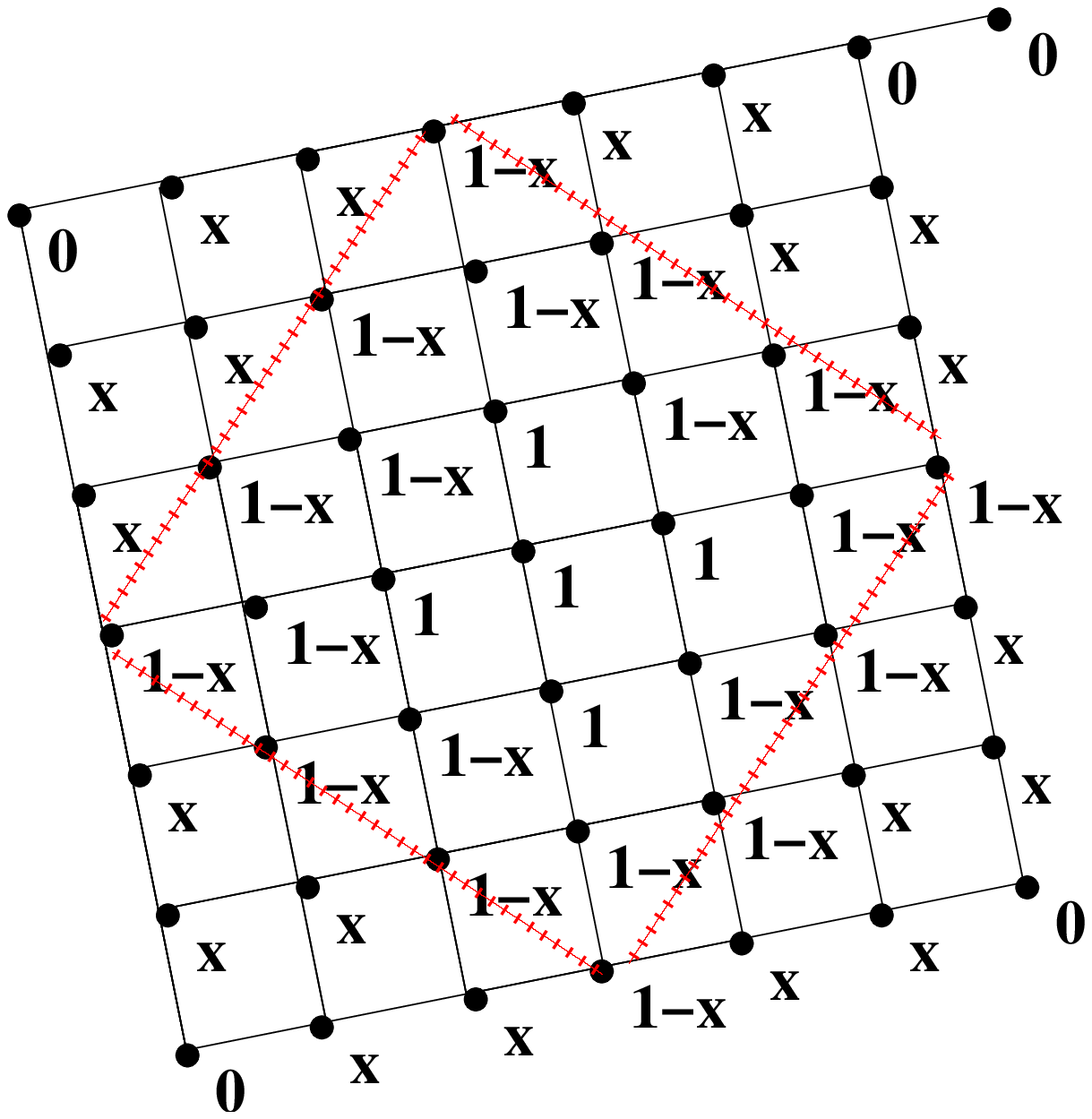}{\vspace*{-0.3cm}}
\caption {(color online) Representation of the function $\varphi_x$ in \textbf{k}-space. The dot line corresponds to the Fermi sea of the non-interacting system.}
\label{fig:kspace}
\end{figure}

 Our starting point will be the Guztwiller wavefunction for non-interacting
electrons. It is defined by $J_C = 1$ and 
\begin{align}
\varphi_\mathbf{k}(\xi_\mathbf{k} < 0) &= 1 \nonumber \\
\varphi_\mathbf{k}( \xi_\mathbf{k} > 0) &= 0
\end{align}
where $\varphi_\mathbf{k}$ is the Fourier transform of
$\varphi(\mathbf{r})$ and $\xi_\mathbf{k} = -2t\,(\cos k_x+\cos k_y) -
\mu_0$ , $\mu_0$ being the chemical potential~\cite{Guztwiller}. It is
the ground state of the tight-binding model where the operator
$\hat{V}$ reduces to a constant. In fact, in one dimension, one can
check, using Eq.~(\ref{potential}) and the Vandermonde determinant
formula that $V_C$ is a constant independent of $C$. It can also be
checked numerically in two dimensions. This type of wavefunction is
very useful because by changing the shape of the Fermi sea, one can
also generate a set of excited states of the tight-binding
model. These are, by definitition, non groundstatable. Having at our
disposal a groundstatable wavefunction and a set of non-groundstatable
wavefunctions, we can ask the question how do we go from one to
another. This can be studied by considering the deformed wavefunction
$|\Psi_x\rangle$ defined by:
\begin{eqnarray}
\varphi_{\mathbf{k}}^x &=& 1 \, \, \, \, 0 \leq |\mathbf{k}| \leq k_1 \nonumber \\
\varphi_{\mathbf{k}}^x &=& 1-x \, \, \, \, k_1 < |\mathbf{k}| \leq k_F  \nonumber \\
\varphi_{\mathbf{k}}^x &=& x \, \, \, \,  k_F<  |\mathbf{k}| < k_2\\
\varphi_{\mathbf{k}}^x &=& 0 \, \, \, \, k_2 \leq  |\mathbf{k}| \leq  \pi \nonumber,
\label{eq:deformedWF}
\end{eqnarray}
$k_F$ being the Fermi momentum. A particular choice of $k_1$ and $k_2$
is represented in \textbf{k}-space on Fig.~\ref{fig:kspace}. By
varying $x$ from $0$ to $1$, we start in the ground state of the
tight binding model and end in an excited state. During the process,
we necessarly lose groundstatability.

\begin{widetext}

\begin{figure}
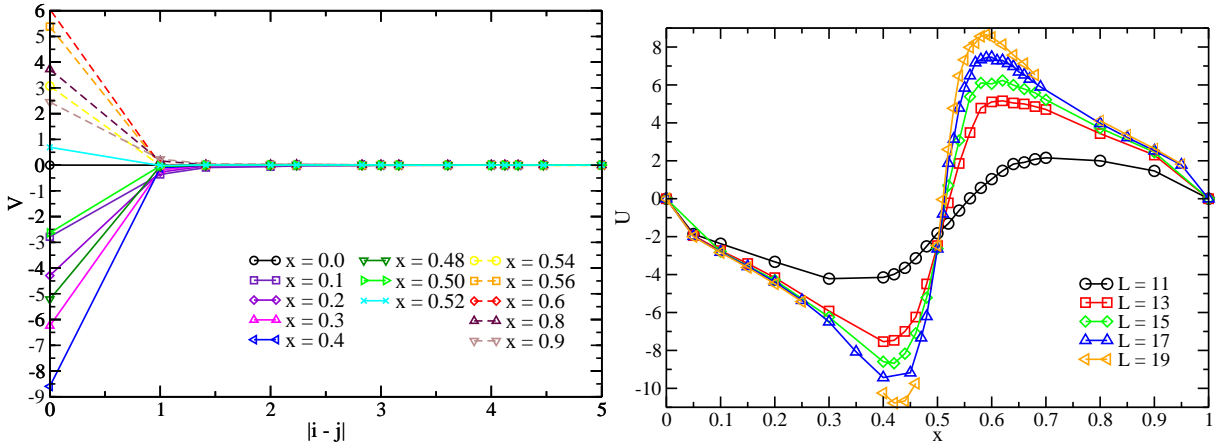

{\vspace*{-0.3cm}}\includegraphics*[width=8cm]{Guztwiller.eps}
{\vspace*{-0.3cm}}\includegraphics*[width=8cm]{Maximum.eps}
\caption {(color online) Left: Evolution of the potential for the
deformed wavefunction (\ref{eq:deformedWF} with $L = 15$ and $ |i - j| < 5$ at
different values of $x$. At large distances, the potential identically
vanishes. Full lines denotes potentials for which the wavefunction is
the groundstate. Dashed lines indicate non groundstatable
wavefunctions. Right: evolution of the Hubbard term $U$ as function of
$x$ for different system sizes.}
\label{fig:Gutz-pot}
\end{figure}

\end{widetext}

\begin{figure}[tr]
{\vspace*{-0.3cm}}\includegraphics*[width=8cm]{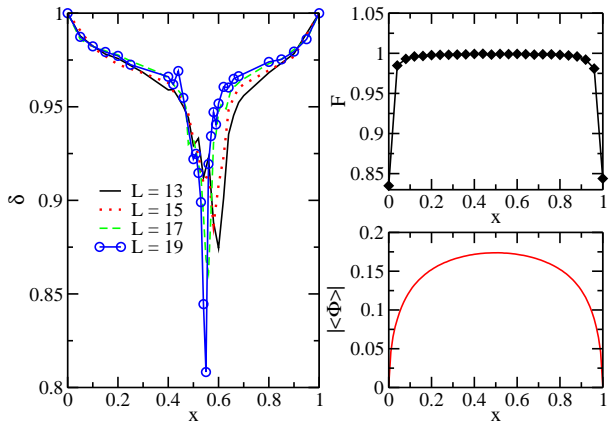}{\vspace*{-0.3cm}}\caption
{(color online) Left: Evolution of $\delta$ as function of $x$ for the
deformed Gutzwiller wavefunction and different system
sizes. Top-Right: evolution of the fidelity $F$ as function of $x$ for
$L = 15$. Bottom-Right: evolution of the superconducting order
parameter $|<\Phi>|$ as function of $x$}
\label{fig:Gutz-ovl}
\end{figure}

We studied the potential obtained from $\varphi_x$ as function of $x$
with the LLS method (see Fig.~\ref{fig:Gutz-pot}). For $x \neq 0$, the
two-body approximation shows a fast-decaying potential. We focused on
the evolution of the Hubbard term $U$ of this potential. As $x$
increases from $0$, $U$ becomes first more and more negative, then
abruptly changes sign around a critical value $x_C \approx 0.55$,
becomes largely positive, and finally steps back to zero. Increasing
the system size, the turnaround of $U$ around $x_C$ becomes more and
more brutal. This behavior is also noticed in the other coefficients
$V_{ij}$. Another interesting feature is observable through the
evolution of the overlap $\delta$ as function of $x$ (see
Fig.~\ref{fig:Gutz-ovl}, left panel). The overlap exhibits a growing
drop around $x_C$, indicating the presence of large $3$-body and
higher order terms in the expansion of $\hat{V}$. In the thermodynamic
limit, the brutal change in the form of the potential should
eventually lead to a singularity in the Hamiltonian as function of
$x$. We interpret the significant change in the nature of the
potential and the fast increase of the correction to the overlap with
system size near $x_C$ as signatures of the boundary of the
groundstatable region. Precisely, for $x < x_C$ the state $|\Psi_x
\rangle$ is indeed the groundstate of the (attractive) Hamiltonian we
are constructing, and for $x > x_C$ it is just an excited state of the
(repulsive) Hamiltonian. 

One can extract additional information on what is happening near $x_C$
by probing the fidelity of the wavefunction \cite{Zou,Zanardi}. The
fidelity, in this context, is a measure of the overlap between two
adjacent states in parameter space:
\begin{equation}
F = | \langle \Psi_x | \Psi_{x + \delta x} \rangle|^2,
\end{equation}
which has been proposed as a useful quantity to expose phase
transitions \cite{Zanardi}. Indeed, at a critical point, $F$ displays
a drop that increases with system size, because the two states
$|\Psi_x \rangle$ and $|\Psi_{x+\delta x} \rangle$ describe two
different phases of matter. In the particular case of a level crossing
(first-order quantum phase transition), the critical point also
corresponds to a loss of groundstatability. 

We computed the evolution of the fidelity for $|\Psi_x \rangle$ with
$\delta x = 0.005$ (Fig.~\ref{fig:Gutz-ovl}, top right). The fidelity
does not display any drop around $x_C$. This fact suggests that the
point $x_C$ cannot be interpreted as a critical point (includying a
first order transition). Instead, the situation appears to be that it
is the Hamiltonian itself that becomes singular at $x_C$ (indeed much
similarly to the simple case of the single spin $S=1/2$ discussed in
the introduction). 

We observe two drops of $F$ at $x = 0$ and $x = 1$.  To understand
this, we measured the superconducting BCS order parameter:
\begin{equation}
| \langle \Phi \rangle | = \frac{1}{N} 
\sqrt{ \sum_{\mathbf{r} \mathbf{r}'} \langle c^{\dag}_{\mathbf{r}'\ups}c^{\dag}_{\mathbf{r}'\downs}c_{\mathbf{r}\ups}c_{\mathbf{r}\downs}  \rangle}
\;.
\label{eq:SC-order-parameter}
\end{equation}
For $0 < x <1$, the system develops superconductivity
(Fig.~\ref{fig:Gutz-ovl}, bottom right). Like the fidelity, the SC
order parameter is unable to detect the loss of groundstatability at
$x_C$; the wavefunction is continuous (again, it is the derived
Hamiltonian that is not) and thus the order parameter derived from
this wavefunction is non-singular at $x_C$. But with our analysis of
$\delta$, we now know that, for $x > x_C$, $| \Psi_x \rangle$ is not
the ground state of the Hamiltonian that we constructed and so we
cannot conclude on the presence of superconducting order in the ground
state.  Finally, note that one could have chosen a different
parametrization for the function $\varphi_x$ and a different final
excited state. For example, we checked several choices of $k_1$ and
$k_2$. The evolution of $U$ and $\delta$ have been found to be similar,
just with different values for $x_C$.  

The analysis of the Guztwiller wavefunction gives us the basic steps
to follow in order to determine if a wavefunction is groundstatable:
start from a wavefunction that is known to be the ground state of a
Hamiltonian with a given kinetic energy, then change continuously a
parameter and observe whether there is some rapid evolution of the
potential and of the overlap. If no such feature appears, then the
potential obtained from the LLS is indeed the potential for which $|
\Psi \rangle$ is the ground state of the Hamiltonian $\hat H_\Psi$.

\subsubsection{BCS  wavefunctions}

Introducing BCS pairing correlations between fermions, one can
consider a superconducting wavefunction with
\begin{equation}
\varphi^{s,d}_\mathbf{k} =\frac{\Delta^{s,d}_\mathbf{k}}{\xi_\mathbf{k}
+\sqrt{\xi_\mathbf{k}^2+{\Delta^{s,d}_\mathbf{k}}^2}} 
\;,
\label{eq:BCS}
\end{equation} where
\begin{eqnarray*}
\Delta^s_\mathbf{k}&=& \Delta \\
\Delta^d_\mathbf{k}&=&\Delta (\cos k_x - \cos k_y),
\end{eqnarray*}
and (as in the Gutzwiller case) $\xi_\mathbf{k} = -2t\,(\cos k_x+\cos
k_y) - \mu_0$. Here $\Delta$ and $\mu$ are parameters (related, but
not equal to the actual gap and chemical potential of the system). At
half filling, we take $\mu = 0$ and vary the parameter $\Delta$. We
then compute $\{U,V_{ij} \}$ and $\delta$ for various system sizes.
Let us first discuss groundstatability in this case. The BCS
wavefunctions can be obtained adiabatically from the Guztwiller
wavefunction starting from $\Delta \rightarrow 0$. By adiabatically we
mean that we did not find any singularity in either the Hamiltonian
extracted from the wavefunction or the overlap $\delta$ going from
this limit to a finite $\Delta$.  This is understandable as we expect
to open a gap by increasing $\Delta$. Starting from a known
groundstate wavefunction, we should remain groundstatable as long as
we do not close the gap. Moreover, in the case of the BCS
wavefunctions, groundstatability is further supported by the fact that
these wavefunctions are the ground states of a mean-field effective
Hamiltonian.

\begin{figure}
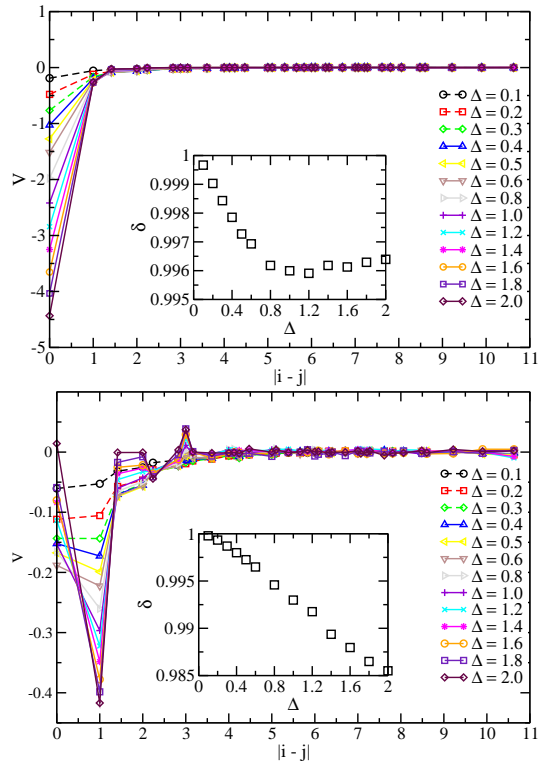

{\vspace*{-0.3cm}}\includegraphics*[width=7cm]{swave.eps}
\includegraphics*[width=7cm]{dwave.eps}{\vspace*{-0.3cm}}
\caption {(color online) Two-body potentials $V_{ij}$ as function of
the distance $|i-j|$ between sites, evaluated at half filling. Top:
the s-wave superconductor. Bottom: d-wave superconductor. Insets: The
overlap is always larger than $98.5 \%$ for the wavefunctions
considered.}
\label{fig:BCSwave}
\end{figure}

The potentials obtained for the $s$-wave state with $L = 15$ are
presented in Fig.~\ref{fig:BCSwave} top. It shows a short distance
two-body negative interaction whose strength is rapidly increasing
with $\Delta$. The potential vanishes when the fermions are separated
by at least three lattice sites. So we see that the $s$ wavefunction
seems a rather good approximation for the attractive Hubbard
model. The evolution of the overlap as function of $\Delta$ is shown
in the Inset. We find $\delta > 0.994$ for $\Delta \leq 2.0$, so the
two-body approximation seems reasonable for these
values. Surprisingly, the overlap seems to converge to a finite value
as function of $\Delta$(!). The two-body potentials for the $d$-wave
case are presented in Fig.~\ref{fig:BCSwave} bottom. The potentials show a
complicate behavior as function of the distance with positive and
negative coefficients in the limit of large $\Delta$. The main common
feature is the presence of a large negative nearest neighbor
interaction term $V_1$, which is consistent with previous mean-field
analysis~\cite{Dagotto}. Additional terms on a longer range are also
non-zero due to the symmetry of the wavefunction.  Comparing the
magnitude of the potentials in the two cases, we find that the
$d$-wave potential is always a lot weaker than the $s$-wave
case. However, the overlap is smaller in the $d$-wave than in the $s$
wave case which means that the two-body approximation is less relevant
for this symmetry.

\subsubsection{Partially projected BCS wavefunctions}

 We also consider the partial Gutzwiller projections of BCS
 wavefunctions. These functions are defined (for the $d$-wave case) by
 $\varphi_{\mathbf{k}} = \varphi^{d}_{\mathbf{k}}$ and a Jastrow
 factor
\begin{equation}
J_C = \prod_i\left(1 - \alpha n_{i\ups} n_{i\downs} \right).
\end{equation}
For $0<\alpha<1$ this factor both penalizes double occupancy and is
positive. Recalling the groundstatability conditions
(\ref{eq:inequal}) obtained from the separable case, we expect that
the wavefunction remains groundstatable as long as the signs in the
wavefunction are not changed. Notice also from (\ref{eq:diagonal})
that a diagonal element of the Hamiltonian becomes ill-defined
(crossing from $\pm\infty$ to $\mp\infty$) if the wavefunction
$\Psi_C$ for a configuration $C$ changes sign. Thus, the
groundstatability of this projected BCS wavefunction is expected from
by the fact that multiplication by a positive Jastrow factor does not
change any signs of the BCS wavefunction. 

\begin{figure}
{\vspace*{-0.3cm}}\includegraphics*[width=7cm]{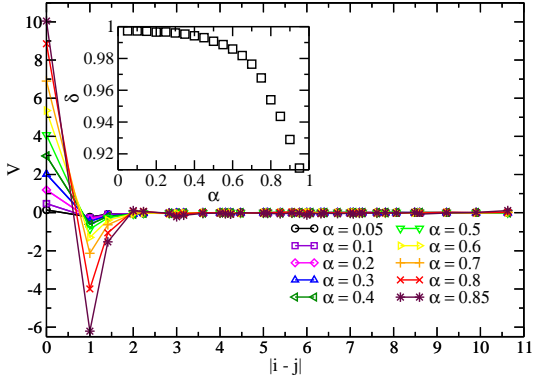}{\vspace*{-0.3cm}}
\caption {(color online) Potential $V_{ij}$ as function of distance
for partially projected BCS wave function with $\Delta_{BCS} =
0.5$. Inset: evolution of the overlap as function of $\alpha$}
\label{fig:Gossamer}
\end{figure}

Figure \ref{fig:Gossamer} presents the evolution of the potential
$V_{ij}$ for different values of $\alpha$ and $\Delta = 0.5$.  The
potential shows a large positive on-site potential growing with
$\alpha$ and a smaller negative short range interaction also growing
with $\alpha$. Thus, for values of $U \sim 10$ relevant for high-$T_c$
superconductivity, we see that a projected BCS wavefunction is favored
by a nearest neighbor attraction of order $V_1 \sim -3$. A Hubbard term
alone is not enough. Moreover, from the evolution of the overlap, we
see that as $\alpha$ gets closer to $1$, the two-body approximation is
less and less justified. In fact, for $\alpha = 1$ the wavefunction
is the resonating valence bond state proposed by Anderson
\cite{Anderson}. A better model to describe such a state would be a
$t-J$ model where interactions are mediated through spin
exchange. This is not allowed in our study since the Heisenberg part
of the $t-J$ model also contains off-diagonal interactions. The point
$\alpha = 1$ is a boundary where the partially projected wavefunction
loses its groundstatability.

\subsubsection{SDW wavefunction}

\begin{figure}
{\vspace*{-0.3cm}}\includegraphics*[width=7cm]{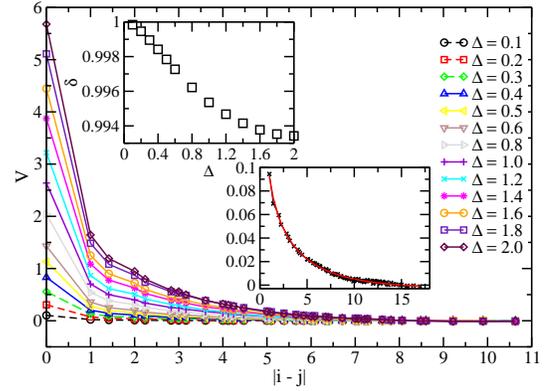}{\vspace*{-0.3cm}}
\caption {(color online) Two-body potentials $V_{ij}$ as function of
the distance $|i-j|$ between sites, evaluated at half filling for the
spin density wave antiferromagnet. Inset: The overlap is always larger
than $99 \%$ The second inset for the SDW wavefunction shows the best
fit for $L = 25$ and $\Delta_{{}_{\rm SDW}} = 0.2$.}
\label{fig:SDwave}
\end{figure}

A spin-density wave state can be generally found in presence of
repulsive interactions, as expected from a mean-field solution of the
Hubbard model. The SDW wavefunction is defined by:
\begin{equation}
\varphi^{{}_{\rm SDW}}(\mathbf{R}_l^{\ups},\mathbf{R}_m^{\downs}) = 
\sum_{\mathbf{k}} \theta(-\xi_\mathbf{k})\;
\alpha_{\mathbf{k}}(\mathbf{R}_l^{\ups})
\;\alpha_{-\mathbf{k}}(\mathbf{R}_l^{\downs})
\end{equation}
 where the sum is restricted to the non interacting Fermi sea and  
 \begin{equation}
 \alpha_{\mathbf{k}}(\mathbf{R}_l^{\sigma}) = u_{\mathbf{k}}e^{i\,\mathbf{k}\cdot\mathbf{R}_l^{\sigma}}+
\sigma\; v_{\mathbf{k}} \,
e^{\,i(\mathbf{k}+\mathbf{Q})\cdot\mathbf{R}_l^{\sigma}}
 \end{equation}
 with $\mathbf{Q} = (\pi,\pi)$ and:
 \begin{eqnarray*}
 u_{\mathbf{k}}^2 &=& \frac{1}{2} \left(1-\frac{\xi_{\mathbf{k}}}{\sqrt{\xi_{\mathbf{k}}^2+\Delta_{{}_{\rm SDW}}^2}} \right) \\
  v_{\mathbf{k}}^2 &=& \frac{1}{2} \left(1+\frac{\xi_{\mathbf{k}}}{\sqrt{\xi_{\mathbf{k}}^2+\Delta_{{}_{\rm SDW}}^2}} \right)
 \end{eqnarray*}
Is the SDW groundstatable? Yes, since we can recover the Gutzwiller
limit by letting $\Delta_{{}_{\rm SDW}} \rightarrow 0$ without
encountering any singularity. The SDW wavefunction should remain
groundstatable for any finite $\Delta_{{}_{\rm SDW}} $. The potential
is presented in Fig.~\ref{fig:SDwave}. It shows a repulsive potential
with a very slow decay. The magnitude of the potential grows with
$\Delta_{{}_{\rm SDW}}$. We find that the potential is well fitted by
an algebraic decay $V(r) \sim 1/r^{\beta}$ with $\beta$ decreasing
when $\Delta_{{}_{\rm SDW}} $ increases. We find $ 0.3 \leq \beta \leq
0.6$ for $0.1 \leq \Delta_{{}_{\rm SDW}} \leq 2.0$. However, it is
very hard to calculate this exponent with accuracy because our system
is not large enough for the potential to really decrease to zero. We
also cannot totally exclude the possibility of a very large but finite
range for the potential. The best fit for a system of size $L=25$ with
$N = 313$ and $\Delta_{{}_{\rm SDW}} = 0.2$ is $ V(r) =
\exp(-r/5.9)/r^{0.4}$ (see second inset of figure~\ref{fig:SDwave}).
Similarly to what we found in the case of the BCS wavefunctions, the
two-body form is a good approximation for the potentials obtained from
SDW wavefunctions as long as $\Delta_{{}_{\rm SDW}}$ is not too large
(the overlap $\delta>99\%$ for $\Delta_{{}_{\rm SDW}}<2$).

 Now, the method is not restricted to these simple cases. Indeed, It
 can be applied to any parametrization of $\varphi(\mathbf{r})$. For
 example, one could search for the Hamiltonian for which a long-range
 wavefunction ({\it e.g.}, with $\varphi(\mathbf{r}) \sim
 1/r^{\alpha}$) is the ground state. The problem here will be not so
 much finding the Hamiltonian but knowing if we are actually starting
 the procedure from the real groundstate or from an excited
 state. Again, we should rely on some adiabaticity argument to answer
 this question. 
  
\subsubsection{Mixed BCS-SDW wavefunction}

\begin{widetext}
 
 \begin{figure}
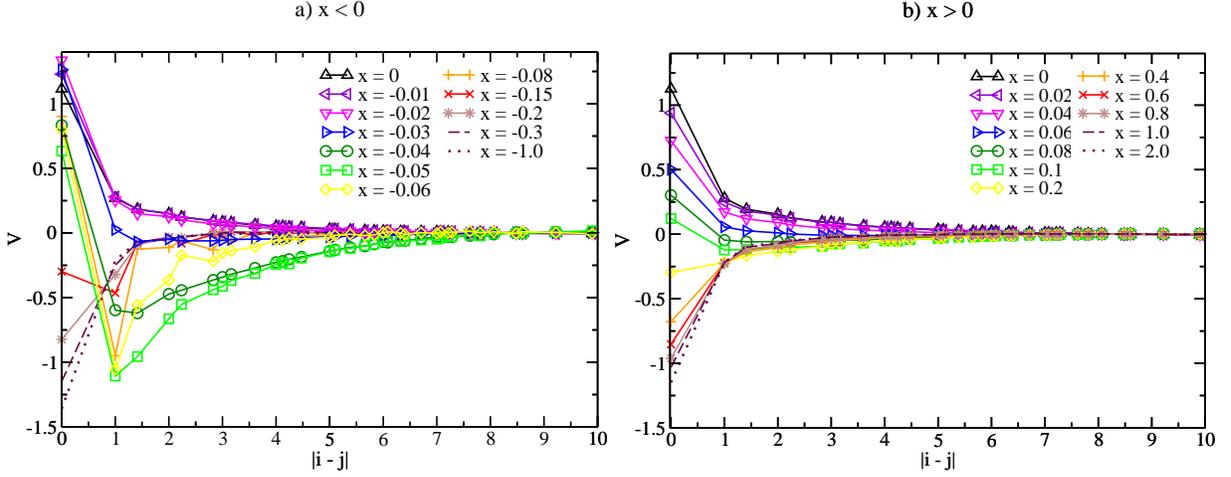

{\vspace*{-0.3cm}}\includegraphics*[width=8cm]{MixL15inf.eps}
{\vspace*{-0.3cm}}\includegraphics*[width=8cm]{MixL15sup.eps}
\caption {(color online) Evolution of the potential as function of $x$ around the pure SDW state} 
\label{fig:Mix}
 \end{figure}
 
 \end{widetext}

Several possibilities exist to construct wavefunctions with both BCS
and AF order \cite{Giamarchi,Sorella,Weber}. These in general rely on
mean-field solutions of Hamiltonians having BCS and SDW
couplings. Here, we will consider a different wavefunction defined by:
\begin{equation}
\varphi_x (\mathbf{R}_l^{\ups},\mathbf{R}_m^{\downs}) =  
x\;\varphi_{{}_{\rm BCS}}^{s} (\mathbf{R}_l^{\ups}-\mathbf{R}_m^{\downs})
+
\varphi_{{}_{\rm SDW}}(\mathbf{R}_l^{\ups},\mathbf{R}_m^{\downs}).
\label{eq:BCS-SDW}
\end{equation}
and we will take $\Delta_{{}_{\rm BCS}} = \Delta_{{}_{\rm SDW}} =
0.5$. Although it is not obvious at first, this form also admits a decomposition in terms 
of single particle wavefunctions, as shown in the appendix.  
We would like to study the groundstatability of this
wavefunction and the evolution of the potential $\hat{V}_x$ as
function of $x$. For very large positive or negative $x$, the
wavefunction reduces to the usual $s$-BCS wave function
(\ref{eq:BCS}). For $x = 0$, this is the pure SDW wavefunction. The
potential we obtain for an arbitrary $x$ is presented in
Fig.~\ref{fig:Mix}. Starting from large negative values of $x$, the
potential is attractive on a short distance. It does not vary much as
soon as $x < -0.3$. When $x$ is approaching zero, the potential
becomes more and more repulsive. The transition is not smooth (see
Fig.~\ref{fig:Mix} left). At some point, the Hamiltonian displays an
attractive long-range potential with a short range repulsion around $x
= -0.06$. This is quite unexpected since the $s$ wave BCS corresponds
to a short range attractive potential and the SDW to a long-range
repulsion. Then, for a short range of very small and negative values
of $x$ ($-0.03 \leq x < 0$), the potential becomes purely repulsive
with a Hubbard term larger than in the pure SDW case, {\it i.e.} $U(x
= -0.02) > U(x = 0)$. Finally, from $x = 0$ to $x$ large and positive,
the potential turns from repulsive to attractive in a very smooth way
(see Fig.~\ref{fig:Mix} right). 

\begin{figure}
{\vspace*{-0.3cm}}\includegraphics*[width=7cm]{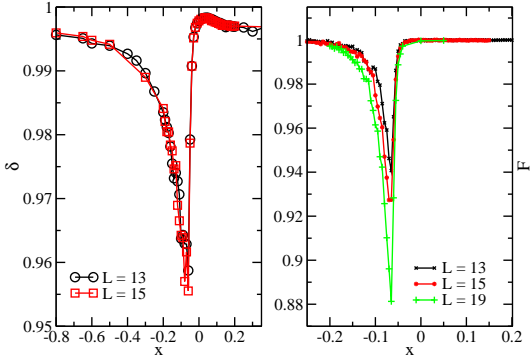}{\vspace*{-0.3cm}}
\caption {(color online) Evolution of the overlap and the fidelity as function of $x$ around the pure SDW state.} 
\label{fig:Ovl2}
 \end{figure}

The evolution of the overlap $\delta$ is shown in Fig.~\ref{fig:Ovl2}
left.  It displays a large drop in the whole region $-0.2 < x <
0$. But it is large and constant for all positive values of $x$ up to
zero.  Another interesting information is given by the fidelity, shown
in Fig.~\ref{fig:Ovl2} right. The fidelity is very close to $1$ for
all positive $x$. In contrast, it displays a sharp drop around $x =
-0.06$ which grows with the system size.

The fact that nothing happens in both the fidelity and the overlap
$\delta$ for $x \geq 0$ leads us to the conclusion that the process of
going from $x = +\infty$ to $x = 0$ preserves the groundstatability of
the wavefunction. Therefore, we are able to find an adiabatic path
between the BCS state and the SDW state. In contrast, for $x = -\infty$ to $x = 0$, we face
a phase transition near $x= -0.06$; this transition is probably
first-order, given 1) the sharpness in the drop of the fidelity, and
2) the fact that the states at $x=\pm\infty$ are the same (they differ
by an overall sign of the wavefunction) and thus cannot be separated
by a second order transition. This particular evolution is peculiar to the mixtured considered in (\ref{eq:BCS-SDW}). For instance,
the wavefunction proposed by Giamarchi and Lhuillier in Ref.~\onlinecite{Giamarchi} does not display this behavior.
 
\begin{figure}
{\vspace*{-0.3cm}}\includegraphics*[width=7cm]{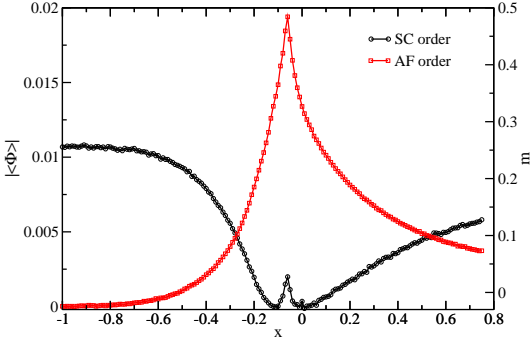}{\vspace*{-0.3cm}}
\caption {(color online) Evolution of the staggered magnetization per site and the SC order parameter for L = 13 as function of $x$. Inset: Evolution of the Fidelity as function of $x$} 
\label{fig:Order}
 \end{figure}

To gain some insights from what happens close to the phase transition,
we also measured the two order parameters: the AF order parameter $m$
defined by
\begin{equation}
m = \frac{1}{N}\sum_{\mathbf{r}}(-1)^{\mathbf{r}}(n_{\mathbf{r} \ups} - n_{\mathbf{r}\downs})
\end{equation} and the SC order parameter (\ref{eq:SC-order-parameter}).
The evolution of the order parameters as function of $x$ are presented
in Figure \ref{fig:Order}. Let us first discuss the evolution of the
AF order parameter. Starting from $x$ large and negative, the
magnetization steadily increases from zero. It then displays a maximum
at $x = -0.06$ and finally decreases back to zero for $x$ large and
positive. So we find that the maximum of the magnetization does not
correspond to the pure SDW state but rather to the SDW state with
small additional BCS correlations. The study of the superconducting
order is also very interesting: $\langle \Phi \rangle$ is maximum for
large values of $|x|$ and it vanishes at $x=0.0$ as expected. But It
also displays an unexpected local maximum at $x=-0.06$. In the range
$-0.12 < x < 0$, both superconductivity and magnetism are not
competing but rather supporting each other. Notice also that the state
with $x=-0.12$ and the pure SDW state share the same characteristics:
they have the same value of $m$ and no BCS order at all. What does
this corresponds to in terms of the Hamiltonian? Turning back to
Fig.~\ref{fig:Mix}, we see that the maximum of both orders corresponds
to a potential with a short range repulsion and a long-range
attraction. However, It is hard to draw a clear conclusion on this
potential since the variation of the overlap $\delta$ is pretty large
near $x=-0.06$. Nonetheless, it appears that we can trust the results
on the potential for $-0.04 \leq x < 0$ where the overlap is still
large. As we already discussed, this corresponds to a purely repulsive
potential with a Hubbard term larger than in the pure SDW case. A more
detailed investigation of this phase could give interesting results on
the possibility of having BCS order with purely repulsive
Hamiltonians.
 
\label{sec:BCS-SDW}
\section{Conclusion}

In conclusion, although the question here posed of groundstatability
of a given wavefunction for a fixed kinetic energy term is much harder
to address in fermionic than in bosonic systems, there are cases where
it can be answered concretely, and we gave examples in this paper.
Considering exact results such as the Calogero-Sutherland wavefunction
in 1D, we were able with our method to recover the Hamiltonian
starting from the wavefunction.  In 2D, we illustrated the problem of
groundstatability on the Guztwiller wavefunction, where by slightly
deforming the Fermi sea of a non-interacting fermion system, one can
go from the groundstate to an excited state.  Then, we analyzed
several mean-field wavefunctions with different types of
superconductivity and antiferromagnetism. These wavefunctions appeared
to be groundstatable as they can be obtained by a deformation of the
Guztwiller state without losing groundstatability in the
process. Starting from these wavefunctions, we obtained the potential
for which these states are the exact ground states, and we showed that
the two-body approximation to the potential appeared to be valid in a
broad range of parameter space. We found both short-range and
long-range interactions in the Hamiltonian. Finally, we discussed the
case of two non-trivial cases: the partially projected Guztwiller
wavefunction and a state with a mixture of superconductivity and
antiferromagnetism. In the latter, we were able to find a Hamiltonian
favoring both SC and AF at the same time.

The approach we follow, of constructing the Hamiltonian starting from
the wavefunction and kinetic term, should be useful in determining
whether certain exotic states of matter -- for example non-Fermi
liquids in two or higher dimensions, and RVB states -- are permitted
in nature. Instead of guessing Hamiltonians that would realize these
states, one can algorithmically determined the target Hamiltonian
starting from a wavefunction and local kinetic terms. Whether such
states of matter exist in nature translates into the question of
whether these wavefunctions are groundstatable or not.

We thank F. Alet and A. Sandvik for enlightening discussions, and
GENCI for allocation of CPU time.  Simulations used the ALPS
libraries~\cite{ALPS}.

\appendix
\section{Building the $\;\det\,[\varphi]\;$ many-body wavefunction from one particle wavefunctions}

It is possible to generalize the results in Refs.~\cite{Gros,Bouchaud}
and write many-body wavefunctions (built by creating particle pairs)
in terms of a determinant of a matrix built from functions
$\varphi(R^{\ups},R^{\downs})$ of two variables, the positions
$R^{\ups}$ and $R^{\downs}$ of up and down particles. It is actually
interesting to ask the reverse question, and find out the conditions
on a function $\varphi(R^{\ups},R^{\downs})$ so that the determinant
of a matrix constructed from this function corresponds to a many-body
wavefunction built by creating particles in pairs. The reason for
addressing this question is that one can then use such function
$\varphi(R^{\ups},R^{\downs})$ to contruct interesting many-body
states where different types of order co-exist.

Consider the state
\begin{equation}
|\Psi\rangle
=
\left(
\sum_\lambda \phi_\lambda\;c^\dagger_{\lambda,\ups}\;c^\dagger_{f(\lambda),\downs}
\right)^N\;
|0\rangle
\end{equation}
where $N$ is the number of pairs, and $|0\rangle$ is the empty
(vacuum) state. The wavefunction is given by
\begin{equation}
\langle\{\mathbf{R}_l^{\ups}\},\{\mathbf{R}_m^{\downs}\}|
\Psi\rangle
=\det [\phi]
\label{eq:det}
\end{equation}
where the $N\times N$ matrix $[\phi]_{lm}\equiv\varphi
\left(\mathbf{R}_l^{\ups},\mathbf{R}_m^{\downs} \right)$, and the
function $\varphi$ is given in terms of the single particle
wavefunctions of the states created by $c^\dagger_{\lambda,\ups}$ and
$c^\dagger_{f(\lambda),\downs}$ and labeled by $\lambda$:

\begin{equation}
\varphi(\mathbf{R}^{\ups},\mathbf{R}^{\downs})
=
\sum_\lambda\;\phi_\lambda\;\a^\ups_{\lambda}(\mathbf{R}^{\ups})\;\a^\downs_{f(\lambda)}(\mathbf{R}^{\downs})
\;.
\label{eq:phiBCS}
\end{equation}

Now, let us suppose that we want to start with a function
$\varphi(\mathbf{R}^{\ups},\mathbf{R}^{\downs})$, and determine how to
decompose it in terms of single particle wavefunctions
$\a^\ups_{\lambda}(\mathbf{R}^{\ups})$ and
$\a^\downs_{f(\lambda)}(\mathbf{R}^{\downs})$ as above. We will do
this construction as follows.

The positions $\mathbf{R}^{\ups}$ and $\mathbf{R}^{\downs}$ take values over the $N_s$
lattice sites $\mathbf{R}_a$, $a=1,\dots,N_s$. We can thus consider the
function $\varphi$ as a $N_s\times N_s$ matrix $\varphi_{ab}\equiv
\varphi(\mathbf{R}^{\ups}=\mathbf{R}_a,\mathbf{R}^{\downs}=\mathbf{R}_b)$. First, note that the matrices
$\varphi^\dag \varphi$ and $\varphi \varphi^\dag$ are hermitian and
thus diagonalizable:
\begin{eqnarray}
\varphi\varphi^\dagger\;\a^\ups_{\lambda}&=&\varepsilon_\lambda\;\a^\ups_{\lambda} \\
\varphi^\dagger\varphi\;\a^\downs_{\lambda}&=&\varepsilon_\lambda\;\a^\downs_{\lambda}.
\end{eqnarray}
That the indices $\lambda$ labeling the states and the eigenvalue are
common can be seen as follows:
\begin{eqnarray}
\varphi^\dagger\;(\varphi\varphi^\dagger)\;\a^\ups_{\lambda}&=&\varepsilon_\lambda\;\varphi^\dagger\;\a^\ups_{\lambda} \\
\varphi^\dagger\varphi\;(\varphi^\dagger\;\a^\ups_{\lambda})&=&\varepsilon_\lambda\;(\varphi^\dagger\;\a^\ups_{\lambda}).
\end{eqnarray}
and therefore $(\varphi^\dagger\;\a^\ups_{\lambda})$ is an eigenstate
of $\varphi^\dagger\varphi$ with eigenvalue
$\varepsilon_\lambda$. Indeed, we can actually pair up the eigenvalues
of $\varphi^\dagger\varphi$ and $\varphi\varphi^\dagger$:
$\a^\downs_{\lambda}\propto(\varphi^\dagger\;\a^\ups_{\lambda})$ and
$\a^\ups_{\lambda}\propto(\varphi\;\a^\downs_{\lambda})$. More
precisely, we can write
$(\varphi^\dagger\;\a^\ups_{\lambda})=\phi^*_\lambda
\a^\downs_{\lambda}$ and $(\varphi\;\a^\downs_{\lambda})=\phi_\lambda
\a^\ups_{\lambda}$, where
$|\phi_\lambda|^2=\varepsilon_\lambda$. (Notice that the phase of
$\phi_\lambda$ can be removed by choosing the overall phase of the
eigenvectors.)

We can thus construct an operator
\begin{equation}
\hat\varphi=\sum_\lambda \phi_\lambda\; 
|\a^\ups_{\lambda}\rangle
\langle  \a^\downs_{\lambda}|
\;,
\end{equation}
from which we can write back the function
$\varphi(\mathbf{R}^{\ups},\mathbf{R}^{\downs})$ by sandwiching $\hat\varphi$ between
$\langle \mathbf{R}^{\ups}|$ and $|\mathbf{R}^{\downs}\rangle$:
\begin{equation}
\varphi(\mathbf{R}^{\ups},\mathbf{R}^{\downs})
=
\sum_\lambda\;\phi_\lambda\;\a^{\ups}_{\lambda}(\mathbf{R}^{\ups})\;
{\a^\downs_{\lambda}}^*(\mathbf{R}^{\downs})
\;.
\label{eq:phi-lambda}
\end{equation}

What we now need is a symmetry 
${\a^\downs_{\lambda}}^*(\mathbf{R}^{\downs})={\a^\downs_{f(\lambda)}}(\mathbf{R}^{\downs})$
that enables us to identify Eq.~(\ref{eq:phiBCS}) with
(\ref{eq:phi-lambda}).

Notice that if $\varphi(\mathbf{R}^{\ups},\mathbf{R}^{\downs})$ is a
real function, and if the spectrum of $\varphi^\dag \varphi$ is non
degenerate, then its eigenvectors $\alpha_{\lambda}^{\downs}$ are
necessarily real and one has
${\a^\downs_{\lambda}}^*(\mathbf{R}^{\downs})={\a^\downs_{\lambda}}(\mathbf{R}^{\downs})$. So
to get a non trivial function $f$, it is important that all
eigenvalues of $\varphi^\dag \varphi$ are degenerate (except maybe at
some particular point where $f(\lambda) = \lambda$), and the
corresponding eigenvectors reside in an eigenspace $E_\lambda$ of
dimension 2. For real $\varphi$ it is possible to write two
orthonormal real eigenvectors in $E_\lambda$, which can then be used
as real and imaginary parts of new orthogonal eigenvectors
$\a^\downs_{\lambda}$ and ${\a^\downs_{\lambda}}^*$. We then identify
the conjugate state ${\a^\downs_{\lambda}}^* \equiv
a^{\downs}_{f(\lambda)}$, where the pair $\lambda,f(\lambda)$ labels
the two states in $E_{\lambda}\equiv E_{f(\lambda)}$, completing the
construction.

What do we need to make the spectrum of $\varphi^\dag \varphi$
degenerate? Suppose the function
$\varphi(\mathbf{R}^{\ups},\mathbf{R}^\downs)$ has some symmetry. For
instance, in the case of the BCS and SDW wavefunctions, one can check
that $ \varphi(\mathbf{R}^\ups,\mathbf{R}^\downs) =
\varphi(-\mathbf{R}^\ups,-\mathbf{R}^\downs)$ is indeed a
symmetry. This symmetry operation, in terms of the matrix $\varphi$,
is implemented through a Hermitian operator $P$ such that $P\varphi
P=\varphi$, or equivalently $[P,\varphi]=0$. It can be trivially
checked that $[P,\varphi^\dag]=0$ as well, and consequently
$[P,\varphi^\dag\varphi]=0$. Because P has two different eigenvalues
$\pm1$, the eigenspaces $E_\lambda$ have dimension 2, which is exactly
what we need to construct the corresponding $f(\lambda)$ to a given
$\lambda$ with $\a^\downs_{f(\lambda)}\equiv {\a^\downs_{\lambda}}^*$.
Therefore, under the conditions above, the identification of
Eq.~(\ref{eq:phiBCS}) with (\ref{eq:phi-lambda}) is complete.

Finally, notice that if one assembles a function from a linear
combination of two functions that satisfy the conditions above (for
example, a symmetry such as $
\varphi(\mathbf{R}^\ups,\mathbf{R}^\downs) =
\varphi(-\mathbf{R}^\ups,-\mathbf{R}^\downs)$ as in the BCS and SDW
cases), then the resulting function also satisfy the conditions. In
particular, the combination
\begin{equation}
\varphi_x (\mathbf{R}_l^{\ups},\mathbf{R}_m^{\downs}) =  
x\;\varphi_{{}_{\rm BCS}}^{s} (\mathbf{R}_l^{\ups}-\mathbf{R}_m^{\downs})
+
\varphi_{{}_{\rm SDW}}(\mathbf{R}_l^{\ups},\mathbf{R}_m^{\downs}).
\end{equation}
does lead to a good fermionic wavefunction (built as in
Eq.~(\ref{eq:det})); this type of wavefunction is the starting point to
the studies that we carried in section~\ref{sec:BCS-SDW}.


\begin{thebibliography}{99}
\bibitem{Castelnovo2005}
C.~Castelnovo, C.~Chamon, C.~Mudry and P.~Pujol, 
Ann.~Phys.~(N.Y.)~\textbf{318}, 316 (2005). 
\bibitem{Arovas-Girvin} D. P. Arovas and S. M. Girvin, in {\it Recent
Progress in Many Body Theories}, edited by T. L. Ainsworth {\it et
al.} (Plenum, New York, 1992), Vol. 3, pp. 315-344.
\bibitem{Marshall} W. Marshall, Proc. Roy. Soc. (London) {\bf A 232}, 48 (1955).
\bibitem{Gros} C. Gros, Phys. Rev. B, \textbf{38}, 931 (1988).
\bibitem{Bouchaud} J.P. Bouchaud, A. Georges and C. Lhuillier,
J. Physique, \textbf{49}, 553 (1988).
\bibitem{Ceperley} D. Ceperley, G.V. Chester and M.H. Kalos, Phys. Rev. B, \textbf{16}, 3081 (1977).
\bibitem{Randeria} A. Paramekanti, M. Randeria and N. Trivedi,
Phys. Rev. B {\bf 70}, 054504 (2004); C. Gros, R. Joynt and T. M. Rice, Z. Phys. B \textbf{68}, 425 (1987) and Phys. Rev. B \textbf{36}, 381 (1987).
\bibitem{Sutherland} B. Sutherland, Phys. Rev. A, \textbf{4}, 2019 (1971)
\bibitem{Guztwiller} M. C. Gutzwiller, Phys. Rev. Lett \textbf{10},
159 (1963); Phys. Rev. \textbf{137}, A1726 (1965).
\bibitem{Zou} W. L. You, Y. W. Li, and S. J. Gu, Phys. Rev. E \textbf{76},
022101 (2007).
\bibitem{Zanardi} P. Zanardi, M. Cozzini and P. Giorda, J. Stat. Mech. (2007) L02002
\bibitem{Dagotto} E. Dagotto \textit{et al.}, Phys. Rev. B,
\textbf{49}, 3548 (1994).
\bibitem{Anderson} P.W. Anderson, Science \textbf{235} 1196 (1987).
\bibitem{Giamarchi} T. Giamarchi and C. Lhuillier. Phys. Rev. B, \textbf{43}, 12943,(1991)
\bibitem{Sorella} L. Spanu, M. Lugas, F. Becca and S. Sorella, Phys. Rev. B, \textbf{77}, 024510 (2008) 
\bibitem{Weber} C\'edric Weber, Andreas Laeuchli, Fr\'ed\'eric Mila, Thierry Giamarchi, Phys. Rev. B \textbf{73}, 014519 (2006) 
\bibitem{ALPS} F. Albuquerque {\it et al.}, J. Magn. Magn. Mater. {\bf 310},
1187 (2007); M. Troyer, B. Ammon and E. Heeb, Lect. Notes Comput. Sci., {\bf
1505}, 191 (1998). See {\tt http://alps.comp-phys.org}.


\end{thebibliography}
\end{document}